\documentclass[amsmath,
 amssymb,
 aps,
 nofootinbib,twocolumn,
 preprintnumbers,
]{revtex4-1}
\usepackage[utf8]{inputenc}
\usepackage{simplewick}
\usepackage{footnote}
\usepackage{graphicx}
\usepackage{dcolumn}
\usepackage{bm}
\usepackage{stmaryrd}
\usepackage{amsmath}
\usepackage{physics}
\usepackage{amssymb}
\usepackage{mathtools}
\usepackage{bbm}
\usepackage{amsfonts}
\usepackage{xcolor}
\usepackage{footmisc}
\usepackage{comment}
\usepackage[normalem]{ulem}
\usepackage{slashed}
\usepackage{multirow}
\usepackage{appendix}
\usepackage{float}
\usepackage{braket}
\usepackage[linktocpage=true,bookmarksnumbered=true,colorlinks=true,plainpages,linkcolor=blue,citecolor=red]{hyperref}
\usepackage{cancel}

\begin{document}
\preprint{ADP-25-T1289, CPPC-2025-08}
\title{ Radion Portal Freeze-Out Dark-Matter}

\author{R. Sekhar Chivukula$^{a}$}
\author{Joshua A. Gill$^{d}$}
\author{Kenn S. Goh$^{b}$}
\author{Kirtimaan A. Mohan$^{c}$}
\author{George Sanamyan$^{b}$}
\author{Dipan~Sengupta$^{d}$}
\author{Elizabeth H. Simmons$^{a}$}
\author{Xing Wang$^{a,e}$}

\affiliation{$^{a}$Department of Physics and Astronomy, University of California, San Diego, 9500 Gilman Drive, La Jolla, CA-92093, USA
}
\affiliation{$^{b}$ARC Centre of Excellence for Dark Matter Particle Physics, Department of Physics, University of Adelaide, South Australia 5005, Australia}
\affiliation{$^{c}$Department of Physics and Astronomy, 
Michigan State University\\
567 Wilson Road, East Lansing, MI-48824, USA
}

\affiliation{$^d$Sydney Consortium for Particle Physics and Cosmology, School of Physics,\\ The University of New South Wales, Sydney NSW 2052, Australia}

\affiliation{$^e$Dipartimento di Matematica e Fisica, Università degli studi Roma Tre,\\ Via della Vasca Navale 84, I-00146, Rome, Italy}

\begin{abstract}
We show that, in a consistent model of a stabilized extra-dimensional theory, the radion can serve as a natural portal between ordinary matter and WIMP dark matter. With an effective coupling scale of the Kaluza-Klein theory of 20$-$100 TeV, the radion portal can produce the observed relic abundance through resonant annihilation for dark matter masses up to a TeV. Existing and planned direct dark matter detection experiments cannot constrain this model. However, indirect detection limits exclude dark matter masses between 5 and 80 GeV, where the radion mediator primarily decays into $b$-quarks.
\end{abstract}

\maketitle


\vfill\eject


Non-baryonic dark matter (DM), which is detected only by its gravitational effects on stars, galaxies, and the cosmic microwave background \cite{Bertone:2016nfn,Cirelli:2024ssz,Planck:2018vyg}, is a major component of the energy density of the Universe.  The WIMP miracle \cite{Kolb:1990vq} suggests that if Weakly Interacting Massive Particles ``froze out" of thermal equilibrium in the early universe, their relic abundance can give rise to today's DM density -- though the extensive results from DM direct detection experiments severely constrain many WIMP models with dark matter masses from a (few) GeV to a TeV \cite{LZCollaboration:2024lux}.\looseness=-1

Extra-dimensional beyond the standard model (BSM) theories, where the weak scale arises from a higher-dimensional gravitational scale \cite{Antoniadis:1990ew,ArkaniHamed:1998rs,Randall:1999ee,Randall:1999vf}, offer a compelling home for WIMP DM. Specifically, in Randall-Sundrum (RS1) theories, 4D spacetime is a slice of a 5D Anti-de Sitter (AdS) space, bounded by a ``Planck brane" and a ``TeV brane". If Standard Model particles (especially the Higgs) are on the TeV brane, the vast difference between the Planck and weak scales is elegantly rephrased in terms of geometry \cite{Randall:1999ee}. In this setup, DM could naturally reside on the TeV brane, interacting with the Standard Model primarily through the extended gravitational interactions.\looseness=-1

Here, we demonstrate that in a consistent model of a stabilized warped extra dimension \cite{Goldberger:1999uk,Goldberger:1999un,DeWolfe:1999cp}, a resonantly annihilating radion -- the lightest scalar excitation of the extra-dimensional gravitational sector -- can act as a natural portal between the Standard Model and WIMP DM. This holds if the effective coupling scale of the Kaluza-Klein (KK) theory \cite{Kaluza:1921tu,Klein:1926tv} is 20-100 TeV. Unlike previous work (see, {\it e.g.}~\cite{Blum:2014jca}), we analyze the radion portal accounting for the significant background metric deformations (``back-reaction") arising from the physics which fixes the size of the extra dimension. The consistency of the model demands the radion mass to be below the TeV scale.\looseness=-1

Intriguingly, while current and planned direct dark matter detection experiments cannot constrain this model (given the coupling scale needed for WIMP relic abundance), indirect constraints from DM annihilation in dwarf spheroidal galaxies impose strong limits when the radion primarily decays into $b$-quarks. This leaves viable DM masses either below 5 GeV or between roughly 80 GeV and 1 TeV. In these allowed regions, the primary signal of this scenario would be the detection of the theory's KK graviton(s).\looseness=-1

The gravitational action for the stabilized RS model (see \cite{Chivukula:2019rij,Chivukula:2020hvi,Chivukula:2021xod,Chivukula:2022tla} for details) includes a 5D Einstein-Hilbert term, as well as a bulk scalar kinetic term and a scalar potential with boundary terms on the branes \cite{Goldberger:1999uk,Goldberger:1999un} chosen to stabilize the background geometry. The $5$D metric for the background is parameterized as~\cite{Chivukula:2021xod,Chivukula:2022tla,Charmousis:1999rg} 
\begin{equation}
 ds^2  = w(x,\varphi) g_{\mu\nu} dx^\mu dx^\nu -r^2_c  v^2(x,\varphi)d\varphi^2~,
    \label{eqn:metricparametrisation}
\end{equation}
in terms of coordinates $(x^\mu,\varphi)$, where  $\varphi \in (-\pi ,+\pi]$  is the coordinate of the orbifolded extra dimension (with $\varphi$ and $-\varphi$ identified) and $r_c$ sets its size. 
Including fluctuations, the $4$D metric is parametrized as 
$ g_{\mu\nu}\equiv \eta_{\mu\nu} + \kappa \hat{h}_{\mu\nu}(x_{\mu},\varphi)$,
where $\eta_{\mu\nu}$ is the usual Lorentz metric (in the``mostly-minus" convention) and $\hat{h}_{\mu\nu}(x,\varphi)$ parametrizes the gravitational fluctuations.\footnote{ $\kappa^{2}=4/M_{5}^{3}$, where $M_{5}$ is the fundamental 5D Planck constant.}. The quantities $w$ and $v$ are defined as 
\begin{align}
    w &\equiv \exp\left[-2\left(A(\varphi)+\frac{ e^{2A(\varphi)}}{2\sqrt{6}} \kappa\,\hat{r}(x,\varphi)\right)\right]~,\ \ \ \\
    v &\equiv   \left(1 + \frac{ e^{2A(\varphi)}}{\sqrt{6}} \kappa\,\hat{r}(x,\varphi)\right)~,
    \label{eq:metricparam2}
\end{align}
where the function $A(\varphi)$ specifies the background geometry (its deviation from AdS form encodes the effect of the stabilization) and $\hat{r}(x,\varphi)$ includes the gravitational scalar modes.\footnote{We work here in unitary gauge, and only include the physical degrees of freedom encoded in $\hat{h}$ and $\hat{r}$.}\looseness=-1

To investigate the radion DM portal in a consistent scalar/gravity theory, we choose the superpotential inspired DeWolfe-Freedman-Gubser-Karch (DFGK) model \cite{DeWolfe:1999cp} for the scalar sector, in which we can obtain analytic solutions to the background scalar and metric fields.\footnote{The DFGK model is briefly summarized in the supplementary material and described in detail in \cite{Chivukula:2024nzt, Chivukula:2021xod,Chivukula:2022tla}.} The solutions involve the background of a bulk scalar field $\phi_{0}(\varphi)$ (taken to be dimensionless by scaling by an appropriate factor of the 5-dimensional Planck scale) and the modified warp factor $A(\varphi)$. These background fields and the modified warp factor are given by \cite{DeWolfe:1999cp,Chivukula:2021xod,Chivukula:2022tla}, 
\begin{align}
    \phi_0(\varphi) &= \phi_{1}e^{-ur_c|\varphi|}~,\ \ \label{eq:backphi} \\
    A(\varphi) &= kr_{c}|\varphi| - \dfrac{1}{48}\phi_{1}^{2}\bigg[1- e^{-2ur_c|\varphi|}\bigg]\ ~,
    \label{eq:APhi}
\end{align}
where $\phi_1$ is the value of the scalar field on the Planck brane and a positive $ur_c$ controls its variation along the extra dimension.\footnote{In principle, $ur_c$ could be negative \cite{DeWolfe:1999cp}, however, we find no solutions in that case with a heavy radion; the sign of $\phi_1$ is unphysical.} This gravity/scalar background reduces to the unstabilized RS1 model, with $A(\varphi)=kr_c|\varphi|$ and $\phi_0(\varphi)$ constant, when either $ur_c,\phi_1 \to 0$. In this limit, $k$ is the usual RS1/AdS curvature which sets the energy scale hierarchy between the Planck and TeV branes (given by $\exp(kr_c)$), the radion is massless, and the additional states from the stabilization sector decouple \cite{Chivukula:2021xod,Chivukula:2022tla}. \looseness=-1

After compactification, the spectrum and wavefunctions of the spin-2 and the spin-0 KK states are obtained by solving the corresponding Sturm-Liouville mode equations for field fluctuations in the extra dimension. These mode equations depend on the metric and scalar background fields given above, and are subject to appropriate boundary conditions.\footnote{The procedure to obtain the spectrum is given in \cite{Chivukula:2022tla,Boos:2005dc,Boos:2012zz}.} We then expand the tensor and scalar fluctuations of the metric in terms of these normal modes as \cite{Chivukula:2022tla}, 
\begin{align}
    \hat{h}_{\mu\nu}(x,\varphi) &=  \dfrac{1}{\sqrt{ r_{c}}} \sum_{n=0}^{+\infty} \hat{h}^{(n)}_{\mu\nu}(x)\, \psi_{n}(\varphi)~,\ \ \  \\
    \hat{r}(x,\varphi) &= \dfrac{1}{\sqrt{ r_{c}}} \sum_{i=0}^{+\infty} \hat{r}^{(i)}(x)\, \gamma_{i}(\varphi)~.
    \label{eq:mode-expansions}
\end{align}
For the spin-2 sector, the $0$-mode represents the massless graviton of Einstein's general relativity, while the $n\geq 1$ states correspond to the graviton KK modes. In the scalar sector, the lightest mode (labeled by $i=0$) corresponds to the radion, while the $i\geq 1$ states correspond to the Goldberger-Wise (GW) scalar states \cite{Goldberger:1999uk,Goldberger:1999un}.  From these solutions, we can compute the self-interactions among the gravitational modes as well as the couplings between the matter sector and the gravitational sector \cite{Chivukula:2024nzt,Chivukula:2021xod,Chivukula:2022tla}. \looseness=-1

To maximize the radion mass, we examine what happens as the geometry in the bulk deviates from AdS, {\it i.e.} as $ur_c$ and/or $\phi_1$ become large. Of particular note is the value of the Ricci curvature near the Planck brane\footnote{AdS corresponds to positive scalar curvature in our mostly minus metric convention.}
\begin{align}
    R(\varphi) &= \frac{1}{r_c^2} \left[20(\partial_\varphi A)^2-8\partial^2_\varphi A\right] \\ \Rightarrow &R(\varphi=0)r_c^2 =  20\left[k r_c-\dfrac{\phi^2_1 ur_c}{24}\right]^2-\dfrac{2}{3}\phi^2_1(ur_c)^2~.
\end{align}
As we increase $\phi_1$, we see that $R(0)$ switches sign -- that is, the space becomes locally de Sitter (and therefore becomes unstable to local fluctuations in energy density \cite{Barrow:1986yf,Padmanabhan:2002ji}) near the Planck brane once
\begin{align}
    \dfrac{\phi^2_1}{24} > \dfrac{kr_c}{ur_c} \left[ 1-\sqrt{\dfrac{4}{5}}\left(\dfrac{ur_c}{kr_c}\right)^{1/2}+~\cdots\right]~.
\label{eq:phi1-max-value}
\end{align}
The hierarchy between the Planck and TeV scales, on the other hand, is determined by
\begin{align}
    A(\pi)-A(0) & = k r_c \pi - \dfrac{1}{48} \phi^2_1 \left(1-e^{-2ur_c\pi}\right)~.
\end{align}
Solving the mode equations numerically, we find that to obtain both a large hierarchy and a heavy radion, one is forced to the limit in which $\phi_1$ saturates the value shown above with $ur_c/kr_c \ll 1$ (as implicitly assumed in Eq.~(\ref{eq:phi1-max-value})), but $ur_c \pi\simeq{\cal O}(1)$.\footnote{If one continues to explore values of $\phi^2_1$ exceeding that in Eq.~(\ref{eq:phi1-max-value}), one finds a region in which $R(0)$ changes sign again, and the geometry is locally AdS near the Planck brane -- however in this case we find the mass of the lightest graviton KK mode exceeds the KK theory's cutoff scale ($m_{1}> \Lambda_{\pi}$), and the effective KK theory breaks down.}

\begin{figure}[t]
\includegraphics[width=0.95\linewidth]{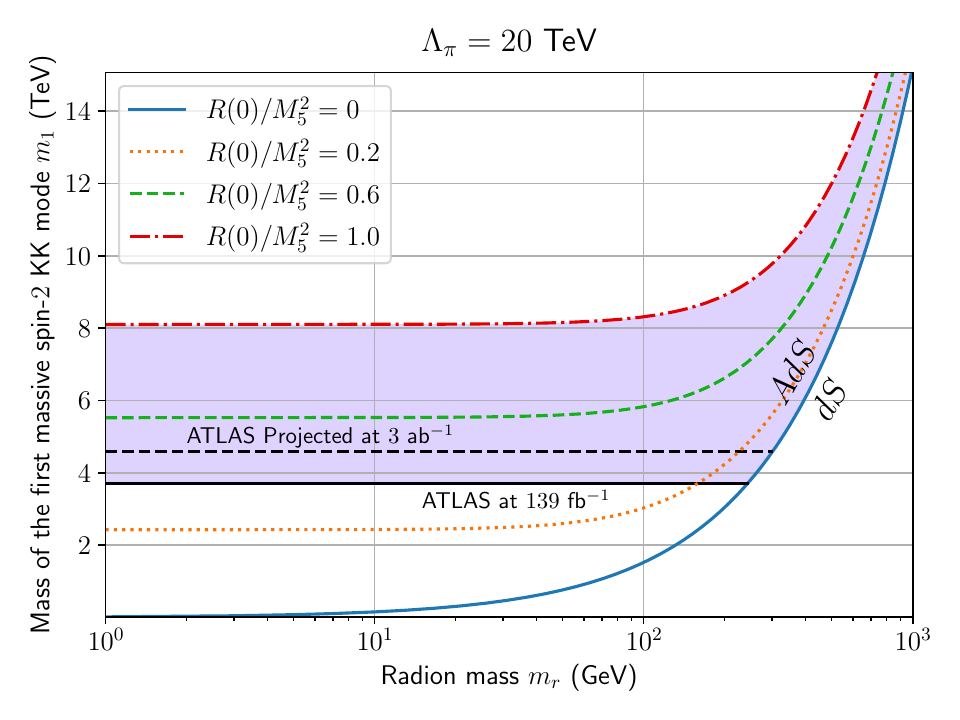}
\caption{Allowed region (shaded) for this model in the $(m_r,m_1)$ plane, for $\Lambda_\pi=20$ TeV. Also plotted are curves of constant $R(0)/M^2_5$, which cannot exceed ${\cal O}(1)$. The horizontal lines are the lower bound on the lightest spin-2 KK graviton \cite{ATLAS:2021uiz}, and projected bounds at the HL-LHC, for this $\Lambda_\pi$. \looseness=-1}
\label{fig:warpfactor1}
\end{figure}

\begin{figure}[t]
\includegraphics[width=0.95\linewidth]{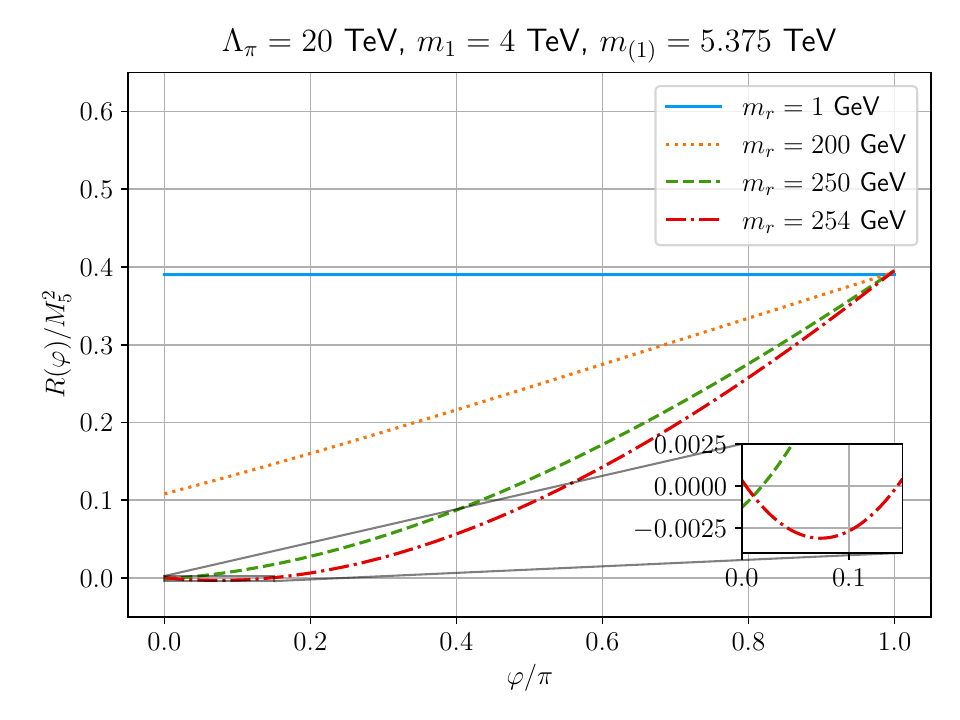}
\caption{Ricci curvature (in units of the five-dimensional Planck scale squared) along extra-dimension for $\Lambda_\pi=20$ TeV and $m_1=4$ TeV, for different radion masses. For radion masses above 250 GeV, the manifold becomes locally de Sitter near the Planck brane.\looseness=-1}
\label{fig:warpfactor2}
\end{figure}

In order to examine the phenomenology of our model, we trade the parameters $(r_c, kr_c, ur_c, \phi_1)$ for the physical parameters of our theory.\footnote{The five-dimensional Planck scale $M_5$ in the Einstein action is set by fixing the four-dimensional Planck scale to its physical value, and this sets all the energy scales of our model.} In the region of interest, where the bound Eq.(\ref{eq:phi1-max-value}) is saturated, the model can be fixed in terms of three quantities: the effective scale of the RS1 theory which determines the couplings of the KK modes $\Lambda_\pi$, the mass of the lowest spin-2 mode $m_1$, and the mass of the radion $m_r=m_{(0)}$.\footnote{In the region of explored here we find $m_r/m_1 \ll 1$, and the GW scalar states are, to a good approximation, just a tower of KK scalar states corresponding to a free massless bulk scalar boson with Neumann boundary conditions on the branes. The scalar masses are then given in terms of zeros of $J_2$, and the ratios of their masses to those of the spin-2 KK states (which are determined by the zeros of $J_1$) are fixed, {\it e.g.} $m_{(1)}/m_1 \equiv 1.34$.}
In Fig.~\ref{fig:warpfactor1}, we present the allowed region of the model in the $(m_r, m_1)$ plane for $\Lambda_\pi=20$ TeV. We include curves of the Ricci curvature at the Planck brane (in units of $M_5$), which must be bounded by ${\cal O}(1)$ in order for our classical gravitational computations to be accurate. Current ATLAS diphoton searches \cite{ATLAS:2021uiz} provide a limit on the KK graviton $m_{1} \ge 4$ TeV \cite{Chivukula:2024nzt}, which implies an upper limit on the radion in our model of 250 GeV for $m_{1}=4$ TeV.  As shown, heavier radion masses are possible for larger values of $m_1$ and/or $\Lambda_\pi$. In Fig.~\ref{fig:warpfactor2} we plot the Ricci curvature (in units of $M_5$) for different radion masses, and the values of the physical parameters shown. Note that, for $\Lambda_\pi=20$ TeV and radion masses above 250 GeV, the back-reaction at the Planck brane is such that the space becomes locally de Sitter.\looseness=-1

Next, we examine the relic abundance of scalar, fermion, or vector DM $\Phi$ confined, along with the Standard Model particles $\psi$, to the TeV brane.
The Lagrangian terms for matter localized on the TeV brane interacting with 
the spin-2 KK and spin-0 GW sector can be found in \cite{Chivukula:2023sua},
and have the usual forms determined by the induced metric on the TeV brane,
$\bar{G}_{\mu\nu} = \left[ wg_{\mu\nu}\right]_{\varphi=\pi}$ and associated covariant derivatives and local vierbeins. The gravitational KK modes couple to the energy-momentum tensor of the TeV-brane localized matter (both DM and SM particles) with a strength proportional to the value of their mode functions evaluated at the TeV brane. This yields, for the lightest massive modes, the couplings 
\begin{align}
    \dfrac{\hat{h}^{(1)}_{\mu\nu}(x)}{\Lambda_\pi}T^{\mu\nu}+\dfrac{1}{\Lambda_\pi} \dfrac{\gamma_0(\pi)}{\psi_1(\pi)}e^{2A(\pi)}\hat{r}^{(0)}(x)T^\mu_\mu~.
\end{align}
Interestingly, in the limit $m_r/m_1 \ll 1$ relevant here, $|\gamma_0(\pi)|e^{2A(\pi)} \approx |\psi_1(\pi)|$ as in  RS1 model with $kr_c \gg 1$.\looseness=-1

\begin{figure}[t]
\includegraphics[width=0.95\linewidth]{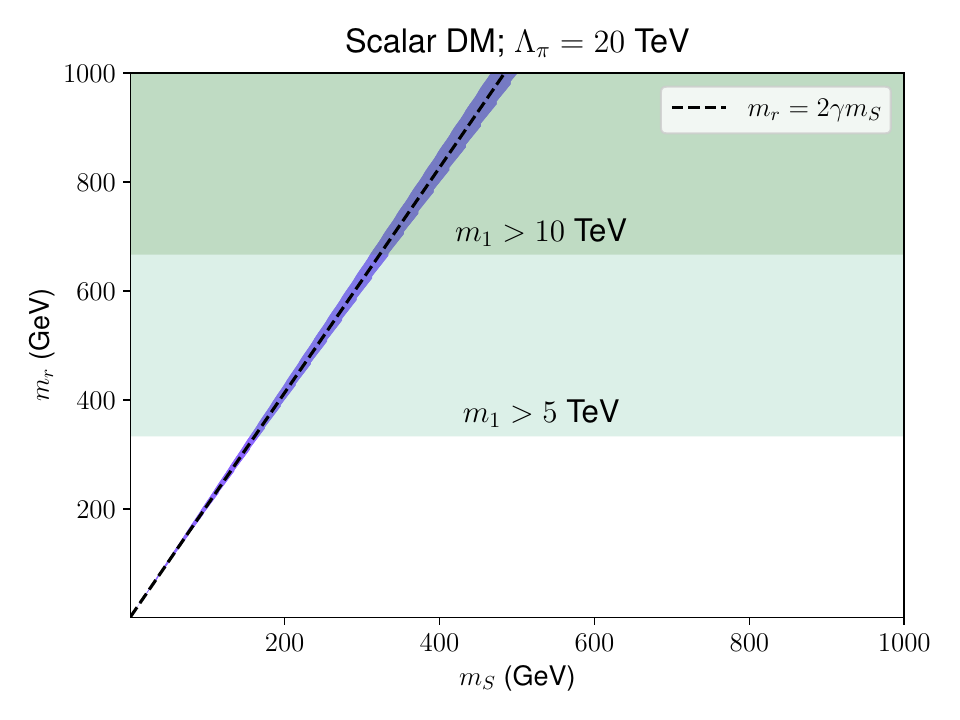}
\caption{Allowed region in $(m_S,m_r)$ space where radion portal scalar DM produces the observed relic density for an interaction strength $\Lambda_\pi=20$ TeV. Note that the parameters cluster around the resonance condition discussed in the text.}
\label{fig:relic}
\end{figure}

To compute the relic freeze-out density, we require the thermal velocity-averaged DM to SM annihilation cross-sections $\braket{\sigma_{\Phi \Phi \rightarrow \bar{\psi} \psi} v}$ \cite{Dodelson:2003ft,Kolb:1990vq}. 
The cross-section proceeds predominantly through $s$-channel radion exchange to all kinematically allowed SM final states. 
We numerically compute the annihilation cross-sections for the various DM candidate types and masses in our model and solve the Boltzmann equation to find the residual density for the allowed regions of the stabilized RS1 model.  The results of this numerical computation are shown for scalar DM in Fig.~\ref{fig:relic}, in which we present the region of parameter space in the $m_{r}-m_{S}$ plane which produces the correct DM relic density for RS scale $\Lambda_{\pi}=20$ TeV. The diagonal dotted line represents the on-resonance annihilation region ($m_{r}=2 \gamma m_{s}$, see below), while the purple region represents the parameter space where the observed Planck-inferred relic DM density is obtained~\cite{Planck:2018vyg}.

We can easily understand our numerical results by making use of the narrow-width approximation for the dominant radion-exchange contribution. Summing over all SM final states, and neglecting the radion branching ratio to DM relative to SM states, we find\footnote{In units such that the Boltzmann constant $k_B$ is one.} 
\begin{equation}
 \braket{\sigma_{\Phi \Phi} v} = \frac{\pi \sqrt{m_r^2 - 4 m_\Phi^2} K_1{\left( \frac{m_r}{T} \right)} \upsilon_\Phi{\left( m_r, m_\Phi \right)}}{48 \Lambda^2_\pi m_\Phi^4 T K_2{\left( \frac{m_\Phi}{T} \right)}^2}~,
 \label{eqn:NWAVelocityAveraged}
\end{equation}
where $K_1{\left( x \right)}$ and $K_2{\left( x \right)}$ are the first and second modified Bessel functions of the second kind and $T$ denotes the temperature of the thermal bath. Here, the interaction-dependent factors
\begin{align}
    \upsilon_{S}{\left( m_r, m_S \right)} &= \left( m_r^2 + 2 m_S^2 \right)^2, \\
    \upsilon_{V}{\left( m_r, m_V \right)} &= \frac{1}{9} \left( m_r^4 - 4 m_r^2 m_V^2 + 12 m_V^4 \right), \\
    \upsilon_{\chi}{\left( m_r, m_\chi \right)} &= \frac{1}{2} m_\chi^2 \left( m_r^2 - 4 m_\chi^2 \right),
\end{align}
differ for scalar $\left( S \right)$, vector $\left( V \right)$, and fermion $\left( \chi \right)$ dark matter candidates.\footnote{Note the helicity-suppression for DM fermions when $m_\chi \simeq 2m_r$.}\looseness=-1

Assuming a standard thermal WIMP freeze-out mechanism, with a typical freeze-out temperature of order $m_\Phi/20$, and therefore the average relative velocity $v=\sqrt{16T/\pi m_\Phi}$,
and the resonant condition $m_\Phi = m_r/2\gamma(v)$ (here $\gamma(v)$ is the Lorentz factor corresponding to the relative velocity $v$), we find the total velocity-averaged cross-sections\footnote{Details of the computation can be found in the supporting material.}
\begin{align}
    \braket{\sigma_{SS} v} \approx \left( 1.7163 \times 10^{-22} \text{ cm}^3/\text{s} \right) \left( \frac{1 \text{ TeV}}{\Lambda_\pi} \right)^2, \\ 
    \braket{\sigma_{VV} v} \approx \left( 6.3807 \times 10^{-24} \text{ cm}^3 / \text{s} \right) \left( \frac{1 \text{ TeV}}{\Lambda_\pi} \right)^2, \\
    \braket{\sigma_{\chi \chi} v} \approx \left( 5.9330 \times 10^{-25} \text{ cm}^3 / \text{s} \right) \left( \frac{1 \text{ TeV}}{\Lambda_\pi} \right)^2. 
\end{align}
Since a velocity-averaged cross-section of $\langle \sigma v_{\rm rel} \rangle\simeq  10^{-26}\, {\rm cm}^{3} / {\rm s}$ can account for the observed observed DM relic density \cite{Dodelson:2003ft,Kolb:1990vq}, this analysis demonstrates a viable resonant scalar (vector) radion portal DM candidate with an effective coupling scale $\Lambda_{\pi}\sim 20-120~(40)$ TeV.\looseness=-1

\begin{figure}[t]
\includegraphics[width=.95\linewidth]{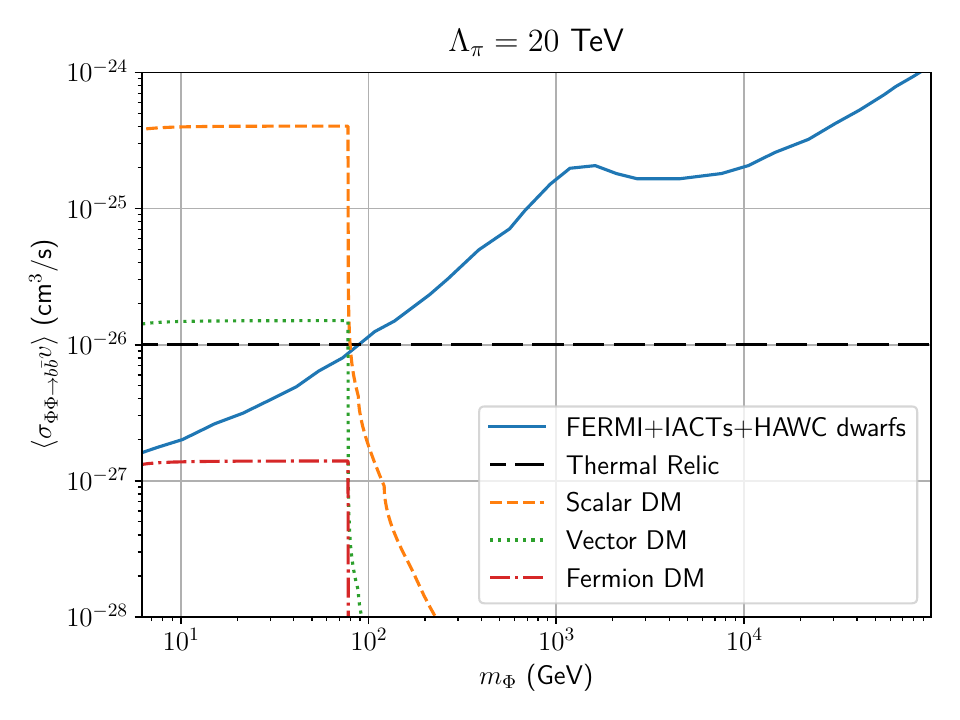}
\caption{ Indirect DM bounds for dark matter annihilating through the radion, which decays entirely to $b$-quarks, plotted against the prediction of the radion-portal model discussed here. DM masses between $\simeq 5-80$ GeV are ruled out \cite{Hess:2021cdp}.}
\label{fig:idd}
\end{figure}

We have also considered the bounds on our radion portal model from DM direct detection and collider searches for the radion \cite{Schumann:2019eaa,LZCollaboration:2024lux,Cooley:2022ufh}, and find them currently too weak to be constraining for $\Lambda_\pi \ge 20$ TeV.\footnote{As illustrated in Fig.~\ref{fig:warpfactor1}, the HL-LHC does have the potential to probe potentially higher KK graviton masses.} Note that for DM masses between $\simeq 5-80$ GeV, the resonant radion mass ($\simeq 2m_\Phi$) kinematically forbids decay into $WW/ZZ$, so the radion decays primarily to $b$-quarks. In this case there are stringent indirect DM bounds \cite{Hess:2021cdp} as shown in Fig.~\ref{fig:idd}. \looseness=-1

We conclude by briefly comparing the radion portal freeze-out discussed here with graviton portal freeze-out as explored in \cite{Chivukula:2024nzt}\footnote{See also \cite{Rueter:2017nbk,deGiorgi:2021xvm}.}. In the latter, the dominant annihilation channel of dark matter to Standard Model particles, which leads to the observed dark matter relic density, occurred in the resonant funnel region where where $2 m_{\rm DM}\simeq m_{\rm KK} $.\footnote{Naively, there are Feynman diagrams with spin-2 and spin-0 final states which would lead to a rapid increase in the velocity averaged cross-section. However, as shown on general grounds in \cite{Chivukula:2019rij,Chivukula:2019zkt,Chivukula:2020hvi,Chivukula:2021xod,Chivukula:2022tla,Chivukula:2022kju, Chivukula:2023sua}, and in the dark matter context in \cite{Chivukula:2024nzt}, there are cancellations between diagrams that lead to a final amplitude where the velocity-averaged cross-section on resonance is the dominant contribution.} Collider constraints push spin-2 KK modes to the multi-TeV scale \cite{ATLAS:2021uiz,Chivukula:2024nzt}. For graviton portal freeze-out, therefore, the DM masses must also be in the multi-TeV range. \looseness=-1

\medskip

{\it RSC and EHS were supported, in part, by the US National Science Foundation under Grant PHY-2210177, and in part at the Aspen Center for Physics, which is supported by National Science Foundation grant  PHY-2210452. KM was supported in part by the US National Science Foundation under Grant PHY-2310497. DS is partially supported by the University of New South Wales, Sydney startup grant PS-71474.
JAG and GS acknowledge the support they have received for their research through the provision of an Australian
Government Research Training Program Scholarship. KSG acknowledges the support he has received for his research through the University of Adelaide Research Scholarship. KSG and GS acknowledge the support provided by the University of
Adelaide and the Australian Research Council through the Centre of Excellence for Dark Matter Particle Physics
(CE200100008). XW is supported by the European Union - Next Generation EU through the MUR PRIN2022 Grant n.202289JEW4. GS is thankful to Prof. Anthony Williams for helpful and fruitful discussions.}

\newpage

\bibliographystyle{apsrev4-1.bst}

\bibliography{main}
\newpage

\clearpage
\onecolumngrid
\pagenumbering{roman}
\setcounter{equation}{0}
\setcounter{figure}{0}
\renewcommand{\theequation}
{\Roman{equation}}
\renewcommand{\figurename}{Supplemental Figure}

\centerline{\bf Supplemental Material for ``Radion Portal Freeze-Out Dark-Matter"}

\subsection{Sturm-Liouville Mode Equations}

\subsubsection{Spin-2}

The 5D tensor fluctuations $h_{\mu\nu}(x,y)$ can be can be decomposed into a tower of 4D KK states $\hat{h}^{(n)}_{\mu\nu}(x)$,
\begin{align}
    \hat{h}_{\mu\nu}(x,\varphi) = \dfrac{1}{\sqrt{ r_{c}}} \sum_{n=0}^{+\infty} \hat{h}^{(n)}_{\mu\nu}(x)\, \psi_{n}(\varphi)~,
    \label{eq:spin2mode-equation}
\end{align}
Here $\psi_n (\varphi)$ is the 5D wavefunction of the $n^{\text{th}}$ mode that satisfies the  SL differential equation:
\begin{align}
    \partial_{\varphi}[e^{-4A}\,\partial_{\varphi}\psi_{n}]=-\mu_{n}^{2} e^{-2A}\psi_{n}.
     \label{eq:spin2-SLeqn}
\end{align}
 These wavefunctions satisfy the Neumann boundary conditions where $(\partial_{\varphi}\psi_{n}) = 0$ at $\varphi \in \{0,\pi\}$. 
The eigenvalues $\mu_n = m_{n} r_{c}$ give the masses $m_{n}$ of the $n^{\text{th}}$ spin-2 KK mode. The wavefunctions are normalized according to 
\begin{align}
    \int_{-\pi}^{+\pi} d\varphi~ e^{-2A}\,\psi_{m}\,\psi_{n} &= \delta_{m,n} \ ,
    \label{eq:spin2-mode-norm}
\end{align}
and satisfy the completeness relation
\begin{align}
    \delta(\varphi_{2}-\varphi_{1}) =  \, e^{-2A}\sum_{j=0}^{+\infty} \psi_{j}(\varphi_{1})\, \psi_{j}(\varphi_{2})~.
    \label{eq:spin2-completeness}
\end{align}
This form of the SL problem for the spin-2 KK sector is {\it identical} to the unstabilized case. The difference is encoded in the new background geometry with a modified warp factor $A(y)$. 

The Neumann boundary conditions $(\partial_{\varphi}\psi_{n}) = 0$ imply that there is always a massless 4D graviton mode (with a wavefunction $\psi_0$ which is constant in $\varphi$) in the spin-2 KK sector. From the form of the spin-2 mode expansion in Eq.~(\ref{eq:spin2mode-equation}) and the constant graviton wavefunction, we immediately find the relationship between the 5D Planck mass and the observed 4D mass $M_{Pl}$
 \begin{equation}
     \dfrac{1}{M^2_{Pl}} = \dfrac{\abs{\psi_0}^2}{r_c M^3_5}~,
 \end{equation}
and hence, using the normalization condition in Eq.~(\ref{eq:spin2-mode-norm}),
\begin{equation}
    M^2_{Pl} = \dfrac{r_c M^3_5}{\abs{\psi_0}^2}=r_c M^3_5\int_{-\pi}^{+\pi} d\varphi\, e^{-2A(\varphi)}~.
\label{eq:4DMPl}
\end{equation}

\subsubsection{Spin-0}

For the spin-0 sector, in which the metric fluctuation and the bulk scalar mix proportional to derivative of background scalar field $\phi^{\prime}_{0} \equiv (\partial_{\varphi}\phi_{0})$,  the KK decomposition  of the 5D scalar field $\hat{r}(x,\varphi)$ into a tower of spin-0 KK modes proceeds by introducing extra-dimensional wavefunctions $\gamma_{i}(\varphi)$ and a tower of 4D scalar fields $\hat{r}^{(i)}(x)$ parameterized as follows:
\begin{align}
    \hat{r}(x,\varphi) = \dfrac{1}{\sqrt{ r_{c}}} \sum_{i=0}^{+\infty} \hat{r}^{(i)}(x)\, \gamma_{i}(\varphi)~.
    \label{eq:spin0-mode-expansion}
\end{align}
The mode equation that brings the 5D scalar Lagrangian to canonical form (including the effects of the mixing between the GW and gravitational sectors) is given by~\cite{Boos:2005dc,Kofman:2004tk},
\begin{align}
    \partial_{\varphi}\bigg[\dfrac{e^{2A}}{(\phi_{0}^{\prime})^{2}} (\partial_{\varphi}\gamma_{i})\bigg] - \dfrac{e^{2A}}{6}\gamma_{i} = -\mu_{(i)}^{2}\,\dfrac{e^{4A}}{(\phi_{0}^{\prime})^{2}} \, \gamma_{i} \, \Bigg\{ 1 + \frac{2\,\delta(\varphi)}{\Big[ 2 \ddot{V}_{1} r_{c} - \frac{\phi _{0}^{\prime\prime}}{\phi_{0}^{\prime}} \Big]} +  \frac{2\,\delta(\varphi - \pi )}{\Big[ 2 \ddot{V}_{2} r_{c} + \frac{\phi _{0}^{\prime\prime}}{\phi_{0}^{\prime}} \Big]} \Bigg\}~,\label{eq:spin0-SLeqn}
\end{align}
where the eigenvalues $\mu_{(n)} = m_{(n)} r_{c}$ give the masses $m_{(n)}$ of the $n^{\text{th}}$ scalar KK mode.  The Dirac delta-function terms enforce the following (orbifold) boundary conditions:
\begin{align}
    (\partial_{\varphi}\gamma_{i})\bigg|_{\varphi = 0} &= -\bigg[2\ddot{V}_{1}r_{c} - \dfrac{\phi_{0}^{\prime\prime}}{\phi_{0}^{\prime}} \bigg]^{-1}\,\mu_{(i)}^{2}\,e^{2A}\,\gamma_{i} \bigg|_{\varphi = 0}~,\nonumber\\
    (\partial_{\varphi}\gamma_{i})\bigg|_{\varphi = \pi} &= +\bigg[2\ddot{V}_{2}\,r_{c} + \dfrac{\phi_{0}^{\prime\prime}}{\phi_{0}^{\prime}}\bigg]^{-1}\,\mu_{(i)}^{2}\,e^{2A}\,\gamma_{i} \bigg|_{\varphi = \pi}~,
    \label{eq:scalarBCtext}
\end{align}
where $\ddot{V}_{1,2}$ are the second functional derivatives of the brane potentials evaluated at the background-field configuration. Note that these boundary conditions reduce to Neumann form in the ``stiff-wall" limit, $\ddot{V}_{1,2} \to +\infty$, a limit which we will use in the DFGK~\cite{DeWolfe:1999cp} method to construct analytic background solutions (see Sec.~\ref{supplement:dfgk} below). The mixing between the gravitational and bulk scalar sectors also generates an
unconventional normalization of the scalar wavefunctions to
bring the scalar kinetic energy terms to canonical form (see \cite{Chivukula:2022tla} and references therein),
\begin{align}
    \delta_{mn} 
    & = 6\int_{-\pi}^{+\pi} d\varphi\hspace{5 pt} \bigg[\dfrac{e^{2A}}{(\phi_{0}^{\prime})^{2}} \gamma_{m}^{\,\prime} \, \gamma_{n}^{\,\prime} + \dfrac{e^{2A}}{6} \gamma_{m}\gamma_{n} \bigg]~. \label{eqScalar:Norm1}
\end{align}
 The scalar wavefunction completeness relation is then given by, 
\begin{equation}
    \delta(\varphi_2-\varphi_1) =  \dfrac{6 e^{4A(\varphi_1)}}{(\phi_{0}^{\prime}(\varphi_1))^{2}} \, \Bigg\{ 1 + \frac{2\,\delta(\varphi_1)}{\Big[ 2 \ddot{V}_{1} r_{c} - \frac{\phi _{0}^{\prime\prime}}{\phi_{0}^{\prime}} \Big]} +  \frac{2\,\delta(\varphi_1 - \pi )}{\Big[ 2 \ddot{V}_{2} r_{c} + \frac{\phi _{0}^{\prime\prime}}{\phi_{0}^{\prime}} \Big]}\Bigg\}~
    \sum_{j=0}^{+\infty} \mu_{(j)}^{2} \gamma_{j}(\varphi_{1})\, \gamma_{j}(\varphi_{2})~.
    \label{eq:spin0-completeness}
\end{equation}
In the scalar sector, due to the non-constant expectation value of the background scalar field, the lightest spin-0 state (identified as the radion with a wavefunction $\gamma_{0}$) acquires a mass $\mu_{(0)}>0$. 

\subsection{DFGK model}
\label{supplement:dfgk}

To find consistent background solutions, we will use the strategy employed in the DFGK model~\cite{DeWolfe:1999cp}, with the introduction of a superpotential-inspired function $W[\hat{\phi}]$ that can be used to derive a GW potential for which the background equations can be easily solved. In this formulation, the scalar bulk and brane potentials are parameterized (in dimensionless form) as:
\begin{align}
    &\hspace{-60 pt}V r_{c}^{2} = \dfrac{1}{8}\bigg(\dfrac{dW}{d\hat{\phi}}\bigg)^{2} - \dfrac{W^2}{24}~,\\
   \varphi\equiv 0: \, \,  V_{1}r_{c} = +\dfrac{W}{2} + \beta_{1}^{2}\bigg[\hat{\phi}(\varphi) -\phi_{1}\bigg]^2~,\hspace{20 pt}&\hspace{20 pt}\varphi\equiv\pi:\, \, V_{2}r_{c} = -\dfrac{W}{2} + \beta_{2}^{2}\bigg[\hat{\phi}(\varphi)-\phi_{2}\bigg]^2~,
   \label{eq:branepotentials}
\end{align}
 and we take the ``stiff-wall" limit: $\beta_{1,2}\to +\infty$, so that $\phi_{1} \equiv \hat{\phi}(0)$ and $\phi_{2} \equiv \hat{\phi}(\pi)$.
 The background scalar and Einstein equations can then be analytically solved to give,
\begin{align}
       (\partial_{\varphi} A) = \frac{W}{12}\bigg|_{\hat{\phi}=\phi_{0}}\text{ sign}(\varphi)~,\hspace{20 pt}&\hspace{20 pt}(\partial_{\varphi} \phi_{0}) = \dfrac{dW}{d\hat{\phi}}\bigg|_{\hat{\phi}=\phi_{0}}\text{ sign}(\varphi)~.\label{AAndPhiInTermsOfW}
\end{align}

The DFGK analysis \cite{DeWolfe:1999cp} then introduces a convenient $W[\hat{\phi}]$ with the following specific form:
\begin{align}
    W[\hat{\phi}(\varphi)] = 12 kr_{c} - \dfrac{1}{2}\hat{\phi}(\varphi)^{2}\,ur_c~.
\end{align}
Plugging this into Eq.~(\ref{AAndPhiInTermsOfW}), we find solutions for the bulk scalar vacuum and the warp factor:
\begin{align}
    \phi_0(\varphi) &= \phi_{1}e^{-ur_c|\varphi|}~, \label{eq:scalarbackgroundi}\\
    A(\varphi) &= kr_{c}|\varphi| + \dfrac{1}{48}\phi_{1}^{2}\bigg[e^{-2ur_c|\varphi|} - 1\bigg]\ ,
    \label{eq:AnPhi}
\end{align}
where the parameters $u$, $\phi_{1}$, and $\phi_{2}$ are related according to
\begin{equation}
    ur_c=\dfrac{1}{\pi}\log\dfrac{\phi_1}{\phi_2}~.
\end{equation}
Given these  $\phi_{0}(\varphi)$ and $A(\varphi)$, we solve numerically for the mass spectrum and wavefunctions of the spin-2 KK sector and the spin-0 GW scalar sector.

\subsection{Physical Parameters}

In order to understand the effect of the back-reaction on the spin-2 masses, we first trade model parameters including the warping scale $k$, the compactification radius $r_{c}$, as well as the bulk scalar potential parameter $u$ and the scalar VEV parameter $\phi_{1}$ for the physical parameters. These include the cut-off scale $\Lambda_{\pi}$, the mass of the first spin-2 KK mode $m_{1}$, the mass of the radion $m_{r}$, and the mass of the first GW mode $m_{\left( 1 \right)}$ as a function of the warp factor $A(\varphi)$. In this case, the parameter space of the model is described by the set of four parameters $\{k, u, r_c, \phi_1 \}$. 
Switching our attention to the output parameters $\{ \Lambda_\pi, m_1, m_{\left( 1 \right)}, m_r \}$, we can express them in terms of the spin-$0$ and spin-$2$ basis functions ($\gamma_n{\left( \varphi \right)}$ and $\psi_n{\left( \varphi \right)}$) as 
\begin{align}
    \Lambda_\pi &= \frac{\psi_0{\left( \pi \right)}}{\psi_1{\left( \pi \right)}} M_{Pl}, \\
    m_1 &= \frac{1}{r_c} \sqrt{ \int_{-\pi}^\pi d \varphi\, e^{-4 A{\left( \varphi \right)}} \left(\psi'_1{\left( \varphi \right)}\right)^2}, \\
    m_{\left( 1 \right)} &= \frac{1}{\sqrt{6} r_c} \left[ \sqrt{ \int_{-\pi}^\pi d \varphi\, \frac{e^{4A{\left( \varphi \right)}}}{\left( \phi'_0{\left( \varphi \right)} \right)^2} \gamma_1{\left( \varphi \right)}^2} \right]^{-1}, \\
    m_r &=  \frac{1}{\sqrt{6} r_c}   \left[ \sqrt{ \int_{-\pi}^\pi d \varphi\, \frac{e^{4 A{\left( \varphi \right)}}}{\left( \phi'_0{\left( \varphi \right)} \right)^2} \gamma_0{\left( \varphi \right)}^2   } \right]^{-1},
\end{align}
where $A{\left( \varphi \right)}$ is the warp factor defined in Eq.~(\ref{eq:APhi}) and $\phi_0{\left( \varphi \right)}$ is the VEV of the bulk scalar field defined in Eq.~(\ref{eq:backphi}). However, as both $A{\left( \varphi \right)}$ and $\phi_0{\left( \varphi \right)}$ are given in terms of the parameters $\{k, u, r_c, \phi_1 \}$ and the basis functions $\gamma_n{\left( \varphi \right)}$ and $\psi_n{\left( \varphi \right)}$ are given by their respective defining differential equations Eq.~(\ref{eq:spin2-SLeqn}) and Eq.~(\ref{eq:spin0-SLeqn}), which are ultimately dependent upon the set of parameters $\{k, u, r_c, \phi_1 \}$ through $A{\left( \varphi \right)}$ and $\phi_0{\left( \varphi \right)}$, we can view the set $\{ \Lambda_\pi, m_1, m_{\left( 1 \right)}, m_r \}$ as a set of implicit functions $\{ \Lambda_\pi{\left( k, u,  r_c, \phi_1 \right)}, m_1{\left( k, u, r_c, \phi_1 \right)}, m_{\left( 1 \right)}{\left( k, u, r_c, \phi_1 \right)}, m_r{\left( k, u, r_c, \phi_1 \right)} \}$.

\par

\subsection{Dark Sector Lagrangian and Couplings}

The Lagrangian for brane-localized scalar fields, vector fields and fermions (both the standard model fields and generic DM candidates) are respectively of the form
\begin{align}
    \mathcal{L}_{S,\text{brane}} &= \int_{-\pi}^{\pi}d\varphi \sqrt{-\bar{G}} \left[ \frac{1}{2} \bar{G}^{\mu\nu} \partial_\mu \hat{S} \partial_\nu \hat{S} - \frac{1}{2} M_S^2 \hat{S}^2 \right] e^{2 A \left( \phi \right)} \delta \left( \phi - \pi \right),\label{eqn:Lphibrane} \\
    \mathcal{L}_{V,\text{brane}} &= \int_{-\pi}^{\pi}d\varphi \sqrt{-\bar{G}} \left[ - \frac{1}{4} \bar{G}^{\mu\rho} \bar{G}^{\nu
    \sigma} F_{\mu \nu} F_{\rho \sigma} + \frac{1}{2} M_{V}^2 \bar{G}^{\mu\nu} \hat{V}_\mu \hat{V}_\nu \right] \delta \left( \phi - \pi \right), \label{eqn:LAbrane} \\
    \mathcal{L}_{\mathcal{\chi}, \text{brane}} &= \int_{-\pi}^{\pi}d\varphi \sqrt{-\bar{G}} \left[ \bar{\hat{\mathcal{\chi}}} i e^{\mu}_{\bar{\alpha}} \gamma^{\bar{\alpha}} D_{\mu} \hat{\mathcal{\chi}} - M_\mathcal{\chi} \bar{\hat{\mathcal{\chi}}} \hat{\mathcal{\chi}} \right] e^{3 A \left( \phi \right)}  \delta \left( \phi - \pi \right)~. \label{eqn:LPsibrane}
\end{align}
In the above, $\bar{G}^{\mu\nu}$ and its determinant are the induced metric on the TeV brane.
\begin{equation}
    \bar{G}_{\mu\nu} = \left[ wg_{\mu\nu}\right]_{\varphi=\pi}~.
\end{equation}
 The vector field strength is $F_{M N} = \nabla_M \hat{V}_N - \nabla_N \hat{V}_M$, and the fermion covariant derivative is defined as 
\begin{equation}
    D_{\mu} \hat{\mathcal{\chi}} = \partial_\mu \hat{\mathcal{\chi}} + \frac{1}{2} \Omega^{\bar{\alpha} \bar{\beta}}_\mu \sigma_{\bar{\alpha} \bar{\beta}} \hat{\mathcal{\chi}}, 
\end{equation}
where $\sigma_{\bar{\alpha} \bar{\beta}} = \left[ \gamma_{\bar{\alpha}}, \gamma_{\bar{\beta}} \right] / 4$, with $\gamma_{\bar{\alpha}}$ being the gamma matrices defined over the tetrad $e^{\mu \bar{\alpha}}$. Without any loss of generality, we take the scalar dark matter candidate to be real, while the fermion is assumed to be Dirac. 
The spin-2 KK modes couple to the energy-momentum tensor of the TeV-brane localized matter through the induced 4D metric on the brane, $\bar{G}_{\mu\nu}=[w g_{\mu\nu}]_{\varphi=\pi}$
\begin{align}
{\cal L}_{\rm spin-2~couplings} =     \dfrac{1}{\sqrt{r_c} M^{3/2}_5} \sum_n {\hat h}^{(n)}_{\mu\nu}(x) \psi_n(\varphi=\pi) T^{\mu\nu} ~,
\end{align}
The scalar fields couple to the trace of the TeV-brane matter energy-momentum tensor,
 \begin{align}
     {\cal L}_{\rm spin-0~couplings} & =\dfrac{e^{2A(\varphi=\pi)}}{\sqrt{r_c} M^{3/2}_5} \sum_i {\hat r}^{(i)}(x) \gamma_i(\varphi=\pi) T^\mu_\mu ~.
 \end{align}

\subsection{Dark Matter Annihilation}

The cross-section for the $2\to2$ annihilation process can be written as 
\begin{equation}
    \sigma_{  \Phi \Phi \rightarrow \bar{\psi} \psi} = \frac{1}{n!} \int \frac{d \Omega}{64 \pi^2} \frac{1}{s} \sqrt{\frac{s - 4 m_\phi^2}{s - 4 m_\Phi^2}} \abs{\mathcal{M}_{  \Phi \Phi  \rightarrow \bar{\psi} \psi}{\left( s , \theta \right)}}^2, \label{eqn:GenericCrossSection} 
\end{equation}
where $\Phi \equiv (S,V,\chi)$ is the dark matter species, $\psi$ is the standard model species that annihilates into the dark matter species, $s$ is the square of the center-of-momentum energy, $n$ is the number of identical particles in the final state, and $\mathcal{M}_{ \Phi \Phi \rightarrow \bar{\psi} \psi}$ is the corresponding $S$-matrix element. The corresponding Feynman diagrams are shown in Supplemental Fig.~\ref{fig:feynman} where the right-hand panel corresponds to DM annihilation to SM via the radion ($n=0$ mode) and Goldberger-Wise modes ($n\geq 1$), the middle panel shows the DM annihilation to spin-0 and spin-2 final states via the radion and GW modes and finally the right-hand panel presents  DM annihilation to radion and GW final states \footnote{The final state radion, GW modes and spin-2 modes finally decay to SM particles.}. In this work radion ($n=0$) portal annihilation to SM particles dominates the matrix elements and the cross-section.

\begin{figure}[t]
$\vcenter{\hbox{\includegraphics[width=0.32\linewidth]{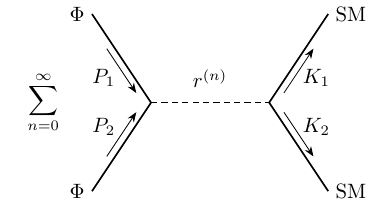}}}+
\vcenter{\hbox{\includegraphics[width=0.32\linewidth]{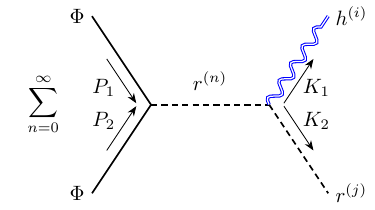}}}+
\vcenter{\hbox{\includegraphics[width=0.32\linewidth]{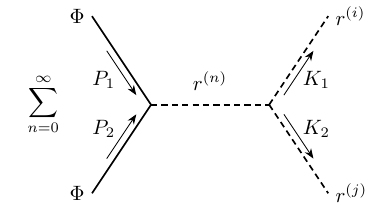}}}$
\caption{Scalar exchange diagrams contributing to DM annihilation: in the region of interest, only the radion ($n=0$) contributes significantly.}
\label{fig:feynman}
\end{figure}

To compute the relic density in the freeze-out setup, we require the velocity-averaged cross-section $\braket{\sigma_{\Phi \Phi \rightarrow \bar{\psi} \psi} v}$, which can be expressed as 
\begin{equation}
    \braket{\sigma_{\Phi \Phi \rightarrow \bar{\psi} \psi} v} = \frac{2 \pi^2 T \int_{4 m_{\Phi}^2}^\infty ds \sqrt{s} \left( s - 4 m_{\Phi}^2 \right) K_1{\left( \frac{\sqrt{s}}{T} \right)} \sigma_{\Phi \Phi \rightarrow \bar{\psi} \psi}{\left( s \right)}}{\left( 4 \pi m_\Phi^2 T K_2{\left( \frac{m_\Phi}{T} \right)} \right)^2}, \label{eqn:VelocityAverage} 
\end{equation}
where $K_1{\left( x \right)}$ and $K_2{\left( x \right)}$ are the first and second modified Bessel functions of the second kind, and $T$ denotes the temperature of the thermal bath. Assuming a standard thermal WIMP freeze-out mechanism, with a typical freeze-out temperature of order $m_\Phi/20$, a velocity-averaged cross-section of $\langle \sigma v_{\rm rel} \rangle\simeq 10^{-26}\, {\rm cm}^{3} / {\rm s}$ can account for the observed Planck-inferred relic density of the Universe~\cite{Planck:2018vyg}. 
We can introduce Eq.~(\ref{eqn:GenericCrossSection}) into Eq.~(\ref{eqn:VelocityAverage}) resulting in an expression relating the velocity averaged cross-section directly to the matrix element 
\begin{equation}
    \braket{\sigma_{\Phi \Phi \rightarrow \bar{\psi} \psi} v} = \frac{1}{n!} \frac{T \int_{4 m_{\Phi}^2}^\infty \int \frac{d \Omega}{32} \frac{ds}{\sqrt{s}} \sqrt{\left( s - 4 m_{\Phi}^2 \right) \left( s - 4 m_\psi^2 \right) } K_1{\left( \frac{\sqrt{s}}{T} \right)} \abs{\mathcal{M}_{\Phi \Phi \rightarrow \bar{\psi} \psi}{\left( s , \theta \right)}}^2}{\left( 4 \pi m_\Phi^2 T K_2{\left( \frac{m_\Phi}{T} \right)} \right)^2}. \label{eqn:SigmaVFinal}
\end{equation}

For the parameter space and the couplings relevant to this work, the decay widths of the radion to the Standard Model, as well as the dark sector, are extremely small, such that we can 
work in the narrow-width approximation
\begin{equation}
    \frac{1}{\left( s - m_r \right)^2 + m_r^2 \Gamma_r^2} \approx \frac{\pi}{m_r \Gamma_r} \delta{\left( s- m_r^2 \right)}, 
\end{equation}
where $m_r$ is the radion mass and $\Gamma_r$ is the radion decay width. We can then write down the $S$-matrix element squared corresponding to the annihilation of various types of dark matter species into the Standard Model species through the radion portal as
\begin{align}
\abs{\mathcal{M}_{\Phi \Phi \rightarrow H H}{\left( s \right)}}^2 &= \frac{16 \pi^2 \kappa_{\left( 0 \right)}^2 }{3} \left[ \sqrt{ 1 - \frac{4 m_H^2}{m_r^2}} \right]^{-1} \upsilon_{\Phi}{\left( m_r, m_\Phi \right)} \beta{\left( r_{\left( 0 \right)} \rightarrow H H \right)} \delta{\left( s -  m_r^2 \right)}, \label{eqn:MatrixElementNWAHiggs} \\
\abs{\mathcal{M}_{\Phi \Phi \rightarrow  Z Z}{\left( s \right)}}^2 &= \frac{16 \pi^2 \kappa_{\left( 0 \right)}^2 }{3} \left[ \sqrt{ 1 - \frac{4 m_{Z}^2}{m_r^2}} \right]^{-1} \upsilon_{\Phi}{\left( m_r, m_\Phi \right)} \beta{\left( r_{\left( 0 \right)} \rightarrow Z Z \right)} \delta{\left( s -  m_r^2 \right)}, \label{eqn:MatrixElementNWAWZ} \\
\abs{\mathcal{M}_{\Phi \Phi \rightarrow W^+ W^- }{\left( s \right)}}^2 &= \frac{8 \pi^2 \kappa_{\left( 0 \right)}^2 }{3} \left[ \sqrt{ 1 - \frac{4 m_{W}^2}{m_r^2}} \right]^{-1} \upsilon_{\Phi}{\left( m_r, m_\Phi \right)} \beta{\left( r_{\left( 0 \right)} \rightarrow W^+ W^- \right)} \delta{\left( s -  m_r^2 \right)}, \label{eqn:MatrixElementNWAWW} \\
\abs{\mathcal{M}_{\Phi \Phi \rightarrow \bar{\psi} \psi}{\left( s \right)}}^2 &= \frac{8 \pi^2 \kappa_{\left( 0 \right)}^2 }{3} \left[ \sqrt{ 1 - \frac{4 m_{\psi}^2}{m_r^2}} \right]^{-1} \upsilon_{\Phi}{\left( m_r, m_\Phi \right)} \beta{\left( r_{\left( 0 \right)} \rightarrow \bar{\psi} \psi \right)} \delta{\left( s -  m_r^2 \right)}, \label{eqn:MatrixElementNWAFermion}
\end{align}
where $\beta{\left( r_{\left( 0 \right)} \rightarrow \psi \psi \right)}$ is the branching ratio corresponding to the radion decay into species $\psi$ as computed from the partial widths quoted below.

\subsection{Spin-$0$ Decay Widths}
\label{sec:decaywidthradion}
We consider the decay of the spin-$0$ KK mode of mass $m_{\left( n \right)}$. We note that the interaction vertex with the brane-localized vector boson is proportional to the mass of the brane-localized vector boson; hence, at the tree level, there will be no contribution to the decay width from decays into $\gamma$ or $g$. We have the following decays into the standard model species localized on the TeV brane
\begin{align}
    \Gamma_{r^{\left( n \right)} \rightarrow H H} &= \frac{\kappa_{\left( n \right)}^2}{192 \pi m_{\left( n \right)}} \sqrt{ 1 - \frac{4 m_H^2}{m_{\left( n \right)}^2} } \left( 2 m_H^2 + m_{\left( n \right)}^2 \right)^2 \label{eqn:GammarHH} , \\
    \Gamma_{r^{\left( n \right)} \rightarrow W W} &= \frac{\kappa_{\left( n \right)}^2}{96 \pi m_{\left( n \right)}} \sqrt{ 1 -  \frac{4 m_W^2}{m_{\left( n \right)}^2}} \left( 12 m_W^4 - 4 m_{\left( n \right)}^2 m_W^2 + m_{\left( n \right)}^4 \right), \label{eqn:GammarWW} \\ 
    \Gamma_{r^{\left( n \right)} \rightarrow Z Z} &= \frac{\kappa_{\left( n \right)}^2}{192 \pi m_{\left( n \right)}} \sqrt{ 1 -  \frac{4 m_Z^2}{m_{\left( n \right)}^2}} \left( 12 m_Z^4 - 4 m_{\left( n \right)}^2 m_Z^2 + m_{\left( n \right)}^4 \right), \label{eqn:GammarZZ} \\
    \Gamma_{r^{\left( n \right)} \rightarrow \bar{f} f} &= \frac{N_c \kappa_{\left( n \right)}^2}{48 \pi} m_f^2 m_{\left( n \right)} \left[ 1 -  \frac{4 m_f^2}{m_{\left( n \right)}^2} \right]^{\frac{3}{2}}, \label{eqn:GammarFF} 
\end{align}
where $f$ is a placeholder for standard model fermions, $N_c$ is the counting factor appearing due to the color charge ($N_c = 3$ for quarks and $N_c = 1$ for leptons), and $m_H$, $m_W$, $m_Z$ are the masses of Higgs, $W$, and $Z$ bosons respectively.

\end{document}